\documentclass[aps,prl,twocolumn,superscriptaddress,showpacs,draft]{revtex4}
\usepackage{graphicx}
\input{epsf}

\begin{document}

\title{Towards Google matrix of brain}
\author{D.L.Shepelyansky}
\affiliation{\mbox{Laboratoire de Physique Th\'eorique (IRSAMC), 
Universit\'e de Toulouse, UPS, F-31062 Toulouse, France}}
\affiliation{\mbox{LPT (IRSAMC), CNRS, F-31062 Toulouse, France}}
\author{O.V.Zhirov}
\affiliation{\mbox{Budker Institute of Nuclear Physics, 
630090 Novosibirsk, Russia}}

\date{February  24, 2010; Revised: April 30, 2010 }


\begin{abstract}
We apply the approach of the Google matrix,
used in computer science and World Wide Web,
to description of properties of neuronal networks.
The Google matrix ${\bf G}$ is
constructed on the basis of neuronal network
of a brain model discussed in PNAS {\bf 105}, 3593 (2008).
We show that the spectrum of eigenvalues
of ${\bf G}$ has a gapless structure
with long living relaxation modes.
The PageRank of the network
becomes delocalized for certain values of the Google damping factor
$\alpha$. The properties of other eigenstates are also analyzed.
We discuss further parallels and similarities 
between the World Wide Web and neuronal networks.
\end{abstract}

\maketitle


\section{I Introduction}

More than 50 years ago John von Neumann traced 
first parallels between architecture of the 
computer  and the brain   \cite{neumann}.
Since that time computers became an unavoidable element of the modern
society forming a computer network connected by the World Wide Web (WWW).
The WWW demonstrates a continuous growth approaching to $10^{11}$
web pages spread all over the world (see e.g. 
 http://www.worldwidewebsize.com/). This number starts to become even
larger than $10^{10}$ neurons in the brain. Each neuron can
be viewed as an independent processing unit connected with 
about $10^4$ other neurons
by synaptic links (see e.g. \cite{izhikevich1,izhikevich2,sporns1}).
About 20\% of these links are unidirectional
\cite{felleman} and hence the brain can be viewed
as a directed network of neuron links.
At present, more and more experimental information
about neurons and their links becomes available and
the investigation of properties of neuronal networks
attracts an active interest of many groups (see e.g.
\cite{laughlin,sporns2,sporns3,kaiser,sporns4,izhikevich3,chklov,monasson}.

The WWW is also a directed network where
a site $j$ points to a site $i$ but no necessary
vice versa. The classification of web sites and information retrieval
from such an enormous data base as the WWW 
becomes a formidable challenge of modern
society where the search engines like Google 
are used by internet users in everyday life.
An efficient way to classify and extract the information
from WWW is based on the PageRank Algorithm (PRA),
proposed by Brin and Page in 1998 \cite{brin},
which forms the basis of the Google search engine.
The PRA is based on the construction of the Google matrix
which can be written as (see e.g. \cite{googlebook} for details):
\begin{equation}
{\bf G}=\alpha {\bf S}+(1-\alpha) {\bf E}/N \; .
\label{eq1}
\end{equation}
Here the matrix ${\bf S}$ is constructed from the adjacency matrix  ${\bf A}$
of  directed network links between $N$ nodes 
so that $S_{ij}=A_{ij}/\sum_k A_{kj}$ and
the elements of columns with
only zero elements are replaced by $1/N$. The  second term
in r.h.s. of (\ref{eq1}) describes  a finite probability $1-\alpha$
for WWW surfer to jump at random to any node so that the matrix elements
$E_{ij}=1$. This term with the Google damping factor $\alpha$ 
stabilizes the convergence of PRA
introducing a gap between the maximal eigenvalue $\lambda=1$
and other eigenvalues $\lambda_i$. As a result
the first eigenvalue has $\lambda_1=1$ and the second one
has $|\lambda_2| \leq \alpha$.
Usually the Google search uses the value
$\alpha=0.85$ \cite{googlebook}. By the construction 
$\sum_i G_{ij}=1$ so that the asymmetric matrix ${\bf G}$
belongs to the class of Perron-Frobenium operators
which naturally appear in the ergodic theory
\cite{sinai} and dynamical systems
with Hamiltonian or dissipative dynamics \cite{mbrin}.

The right eigenvector at $\lambda=1$ is 
the PageRank vector with 
positive elements $p_j$ and $\sum_j p_j=1$.
The classification of nodes in the decreasing order of
$p_j$ values is used to classify
importance of WWW nodes as it is described in more detail in
\cite{googlebook}.
The PageRank can be efficiently obtained by a
multiplication of a random vector by ${\bf G}$
which is of low cost since in average there are
only about ten nonzero elements in a typical line
of ${\bf G}$ of WWW. This procedure converges
rapidly to the PageRank.

Fundamental investigations of the PageRank properties of the WWW
have been performed in the computer science community
(see e.g. \cite{donato,boldi,avrach1,avrach2,litvak,avrach3};
involvement of physicists is visible, e.g. \cite{fortunato1},
but less pronounced).
It was established that the PageRank is satisfactory characterized by 
an algebraic decay $p_j \sim 1/j^\beta$ with
$j$ being the ordering index and 
$\beta \approx 0.9$; 
the number of nodes with the PageRank $p$ scales as 
$N_n \sim 1/p^\nu$ with the numerical value of the exponent
$\nu =1+1/\beta \approx 2.1$ \cite{googlebook,donato}.
It is known that such type of algebraic dependencies appear in various
types of scale-free networks \cite{dorogovtsev}.
The PageRank classification finds its applications not only for
the WWW but also for the network of article citations
in Physical Review
as it is described in \cite{redner,fortunato2}.
This shows that the approach based on the Google matrix can be 
applied to vary different types of networks.

In this work we construct the Google matrix ${\bf G}$ for a model 
of brain analyzed in \cite{izhikevich3}. 
The properties of the spectrum and the eigenstates of ${\bf G}$
are described in the next Section II. 
The results are discussed in Section III.

\section{II Numerical results}

\begin{figure}
\centerline{\epsfxsize=8.5cm\epsffile{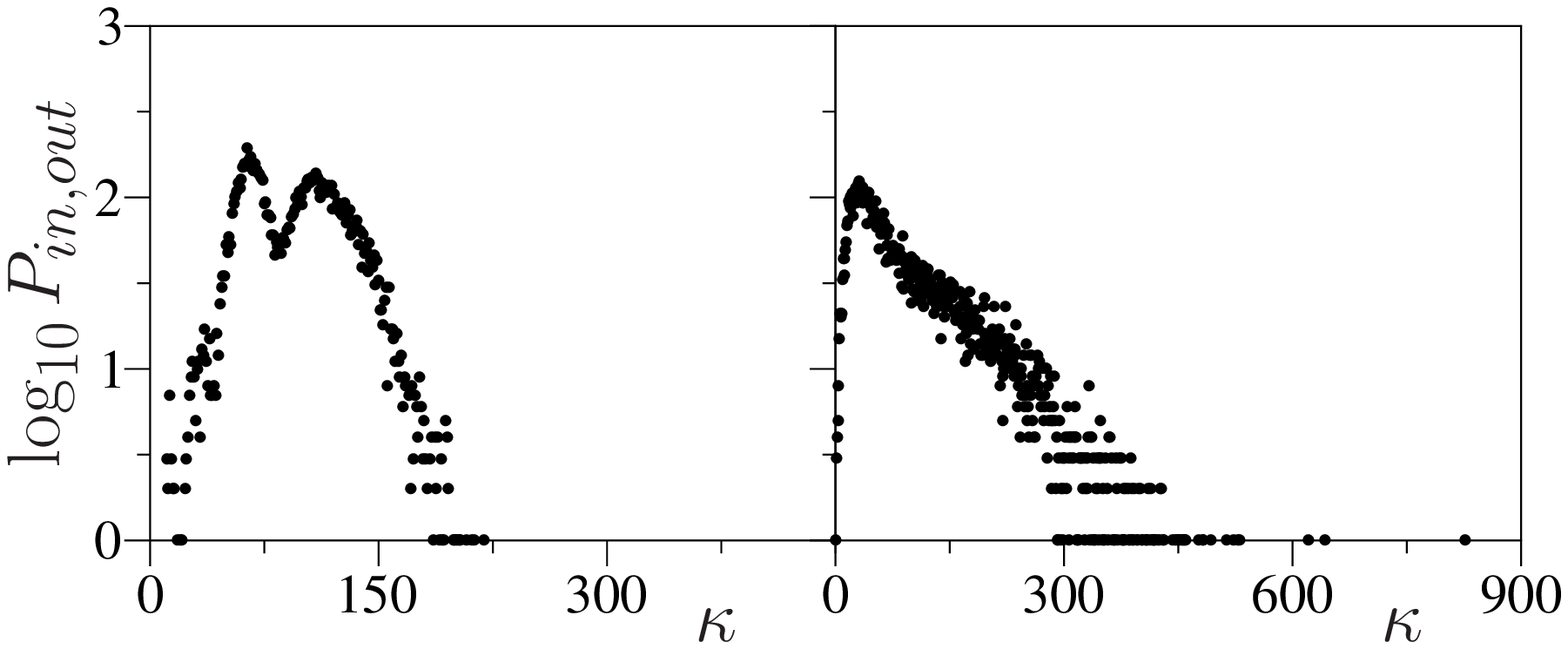}}
\vglue -1.0cm
\centerline{\epsfxsize=8.5cm\epsffile{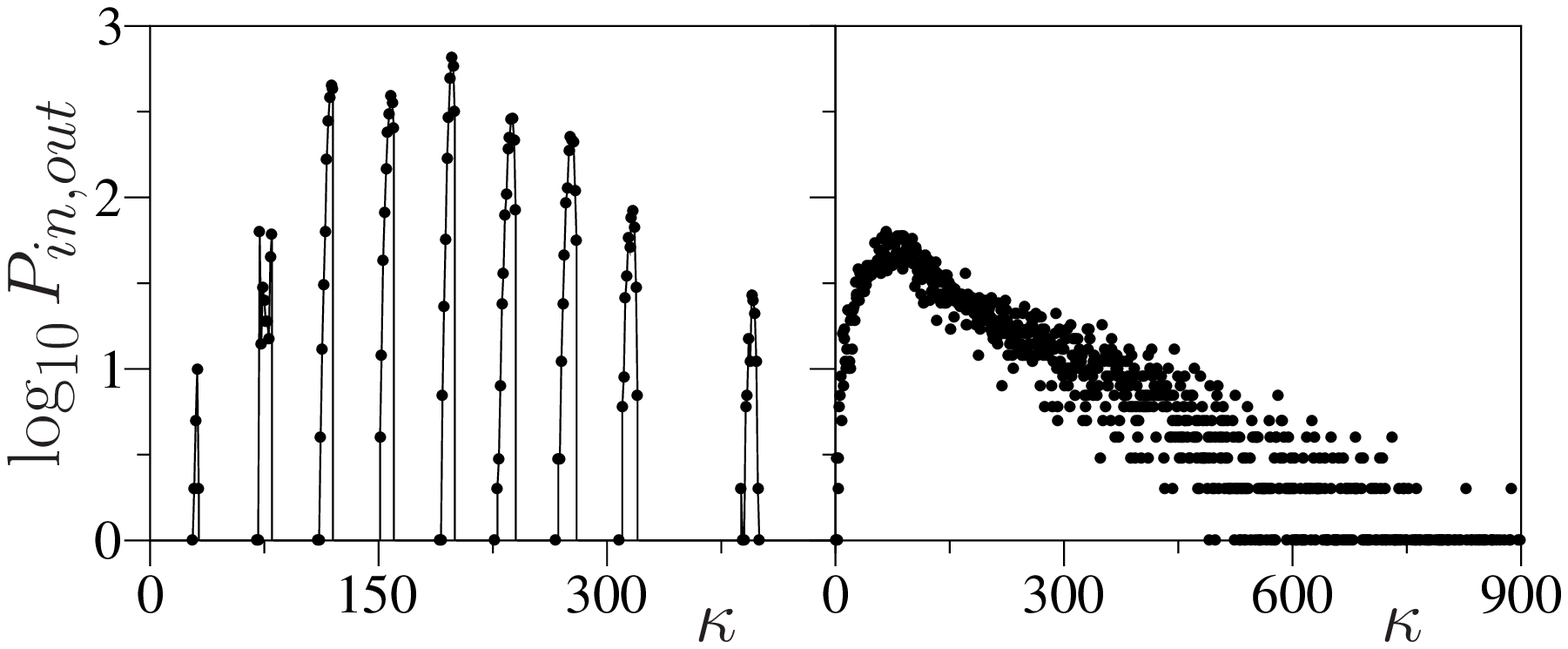}}
\vglue -0.3cm
\caption{Distribution of {\it ingoing} (left panels) 
and {\it outgoing} (right panels) links $\kappa$: 
$P_{in}$ and $P_{out}$ give number of nodes with
$\kappa$ {ingoing} and {\it outgoing} links respectively. 
Top panels: unweighted links; bottom panels: weighted links. 
} 
\label{fig1}
\end{figure}
To construct the Google matrix of brain
we use a directed network of links between $N=10^4$
neurons \cite{izhikevich4}
generated from the brain model \cite{izhikevich3}.
In total there are $N_l=1960108$ links in the network.
They form  $N_{out}$
{\it outgoing} links and $N_{in}$ {\it ingoing} links
($N_l=N_{out}=N_{in}$),
so that there are about 200 outlinks (or ingoing) per neuron.
These numbers include 
multiple links between certain pairs of neurons;
certain neurons have also links to themselves
(there is one neuron linked only to itself).
The number of weighted symmetric links
is approximately $9.8$\%.
Due to existence of multiple links between the same neurons
we constructed two ${\bf G}$ matrices
based on unweighted and weighted counting of links.
In the first case all links from neuron
$j$ to neuron $i$ are counted as one link,
in the second case the weight of the link
is proportional to the number of links from $j$ to $i$.
In both cases the sum of elements in one column 
is normalized to unity.
The distributions of ingoing ($P_{in}$) and outgoing 
($P_{out}$) links 
are shown in Fig.~\ref{fig1}.
The weighted distribution of ingoing links have 
a pronounced peaked structure  corresponding to
different regions of brain model 
considered in \cite{izhikevich3}.
We note that the distribution of links is
not of free-scale type.
\begin{figure}
\centerline{
\epsfxsize=4.5cm\epsffile{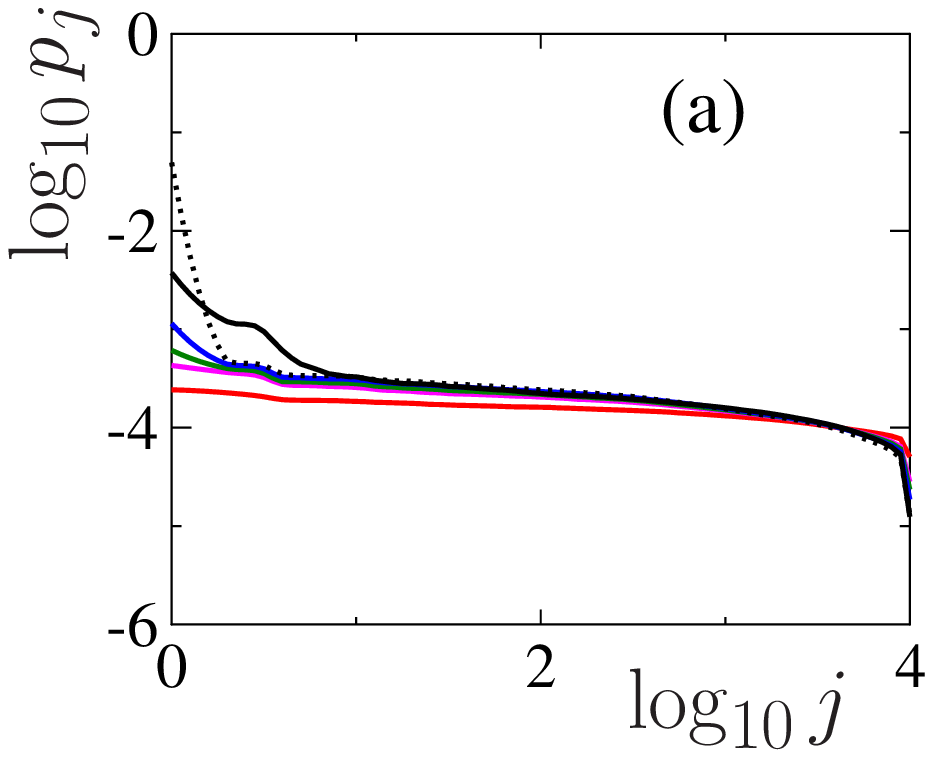}
\epsfxsize=4.5cm\epsffile{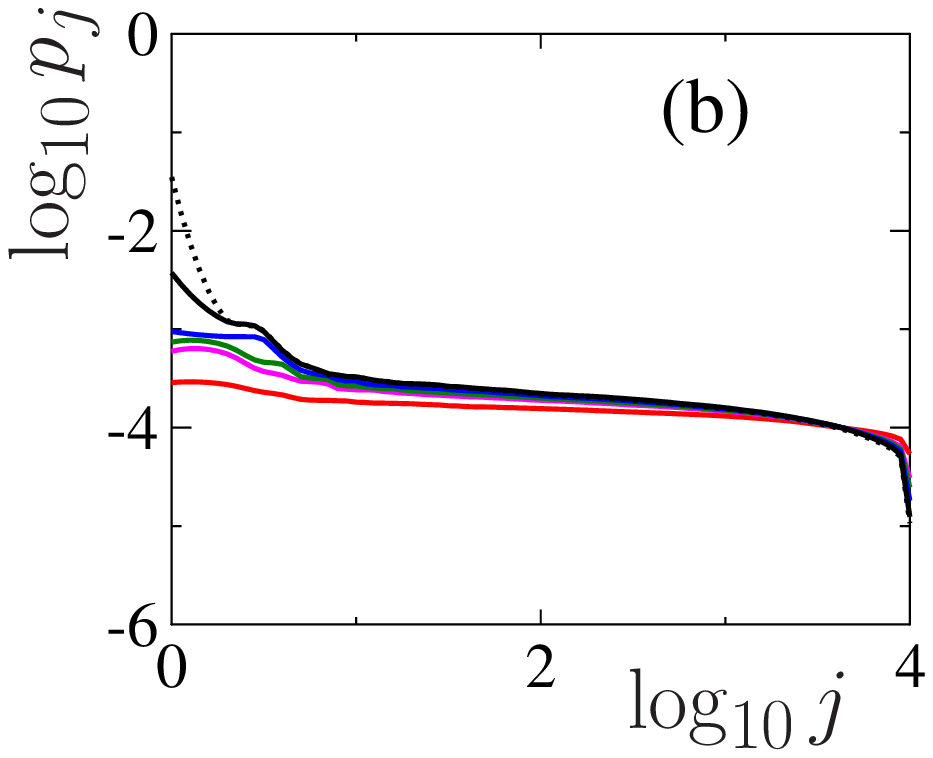}
}
\vglue -0.2cm
\centerline{
\epsfxsize=4.5cm\epsffile{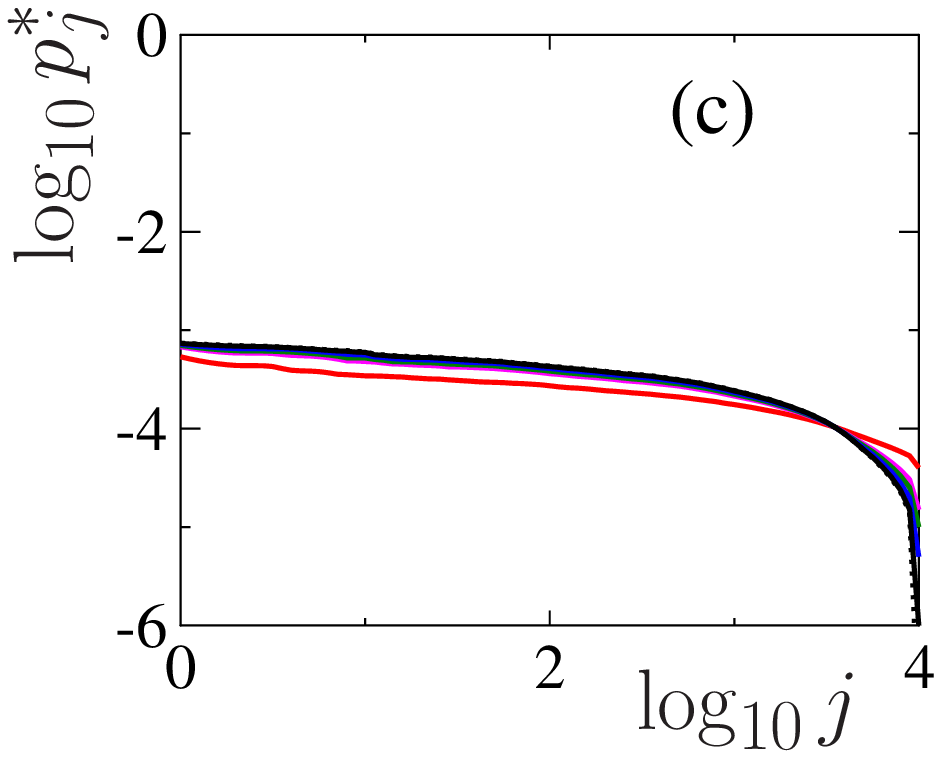}
\epsfxsize=4.5cm\epsffile{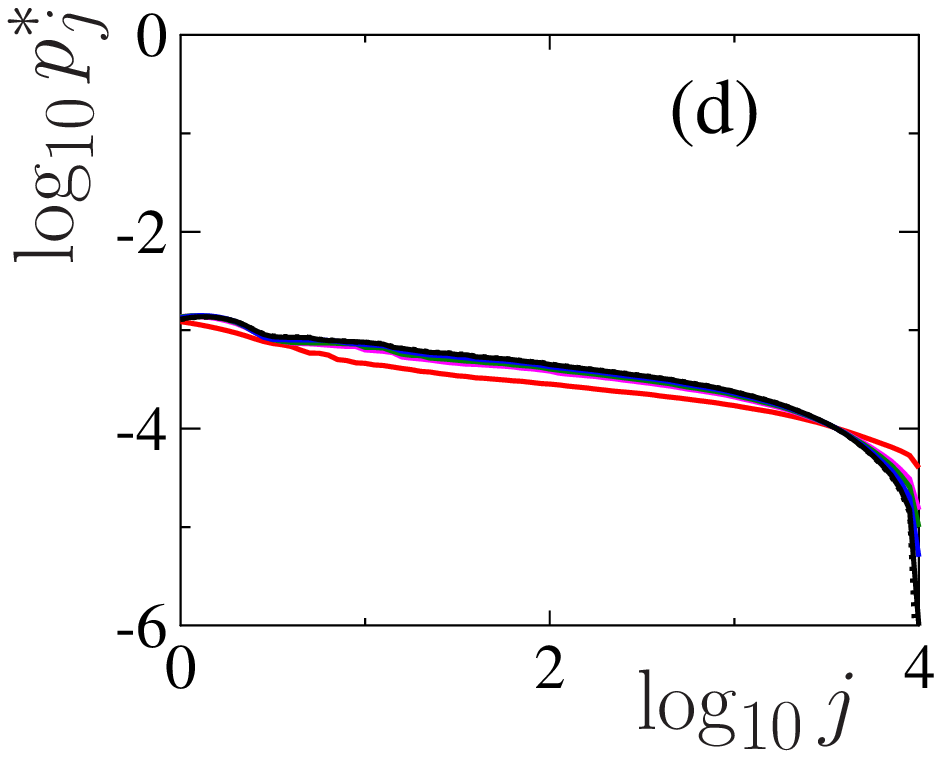}
}
\vglue -0.3cm
\caption{(Color online) PageRank $p_j$ for the Google matrix of brain model
at $\alpha=0.6,0.85, 0.9, 0.95$ and $0.99$ shown by red, magenta
green, blue and black solid curves
(full curves from bottom to top at $\log_{10} j=0.3$). 
The dotted black curve corresponds
to $\alpha=0.999$ and demonstrates strong dependence 
of the PageRank on $\alpha$ in the
vicinity of $\alpha=1$. Panels (a) and (b) 
correspond to unweighted and weighted links.
For panels (a) and (b) 
the values of PAR are $\xi = 8223.$ and $8314.$, $6295.$ and
$6040.$, $5570.$ and $5046.$, $3283.$ and $3367.$, $28.4 $, 
$90.0$, $1.09$ and $1.19$
for $\alpha=0.6, 0.85, 0.9, 0.95, 0.99, 0.999$
respectively. Panels (c) and (d) show the dependence of 
the influence-PageRank $p^*(j)$ on $j$ for the same values
of $\alpha$ as for top panels respectively for 
unweighted and weighted links (for $\alpha >0.6$ there is a
strong overlap of curves).
} 
\label{fig2}
\end{figure}

The dependence of the PageRank on $\alpha$ is shown in Fig.~\ref{fig2}.
For $\alpha=0.999$ almost all probability $p_j$ is concentrated on 
one neuron. This is the only one neuron which is linked
only to itself. With the increase of $\alpha$ up to $ 0.99$
the main part of probability  is concentrated mainly on about 10 neurons
that approximately corresponds to the number of peaks in 
the distribution of weighted ingoing links in Fig.~\ref{fig1} 
(bottom left panel). At the same time 
the PageRank has a long tail at large $j$ where
the probability $p_j$ is practically homogeneous.
For $\alpha=0.6$ the peak of probability
at $1 \leq j \leq 10$ is washed out and the PageRank
becomes completely delocalized.
We note that a delocalization of the PageRank with $\alpha$
appears in the Ulam networks describing dynamical 
systems with dissipation \cite{sz,ermann}.
At the same time the WWW networks
remain stable in respect to variation of $\alpha$ 
as it is discussed in \cite{avrach3,ggs2010}.

Recently, for the studies of procedure call network
of the Linux Kernel \cite{linux}
it was proposed to study the properties of the 
importance-PageRank $p^*(j)$ which is 
given by the eigenvector at $\lambda=1$
for  the Google matrix constructed from the
inverted links of the original adjacency matrix.
It was argued that $p^*(j)$ can give an additional
information about certain important nodes.
Our results for $p^*(j)$ are shown in panels (c,d)
of Fig.~\ref{fig2}. They show that $p^*(j)$
is practically delocalized and flat for all used values of $\alpha$.
This indicates that all nodes have practically equal importance.
The popularity-importance correlator introduced
in \cite{linux} and defined as $\kappa=N \sum_i p(i)p^*(i) -1$
is rather small ($\kappa \approx -0.009, -0.017$ 
at $\alpha=0.6$ and  $\kappa \approx -0.054, -0.065$ at $\alpha=0.85$
for unweighted, weighted links respectively).
This shows that there are no correlations between $p$ and $p^*$
in our neuronal network that is similar to the  
Linux Kernel case.
\begin{figure}
\centerline{
\epsfxsize=4.5cm\epsffile{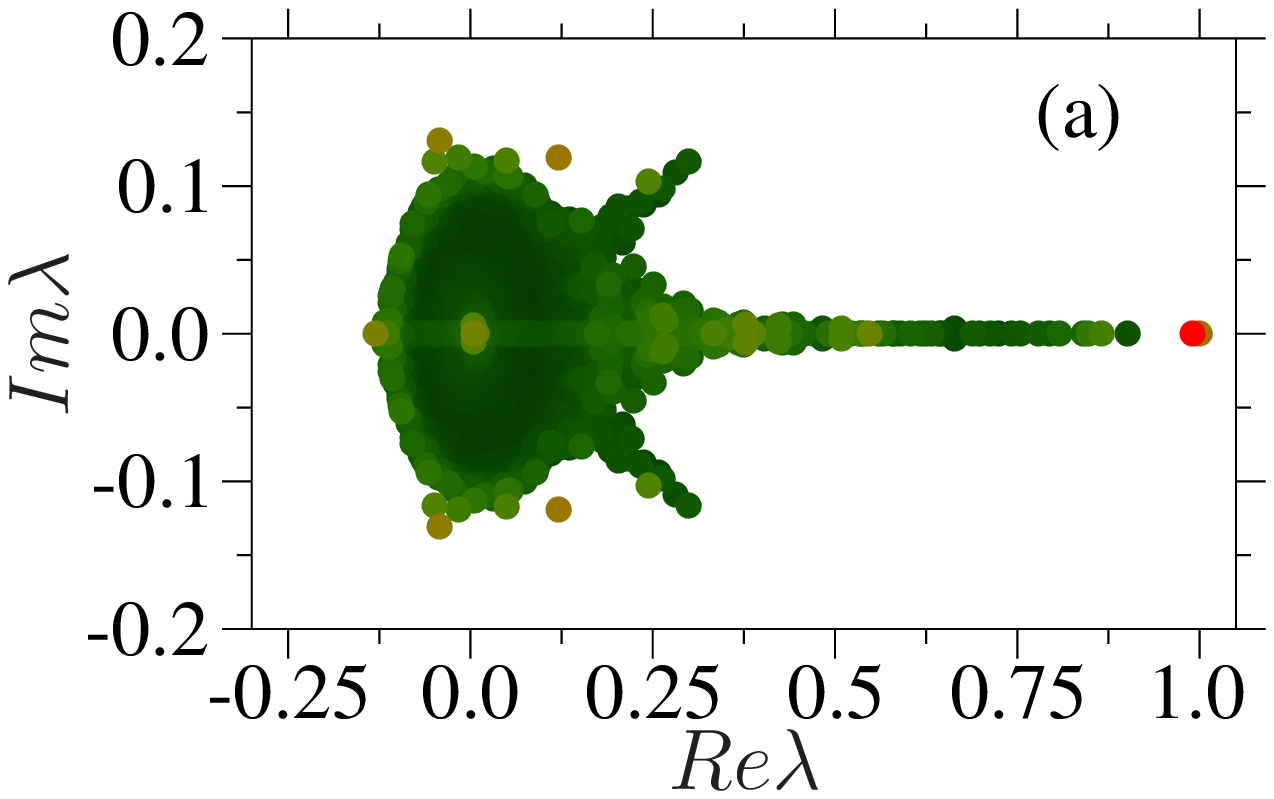}
\epsfxsize=4.5cm\epsffile{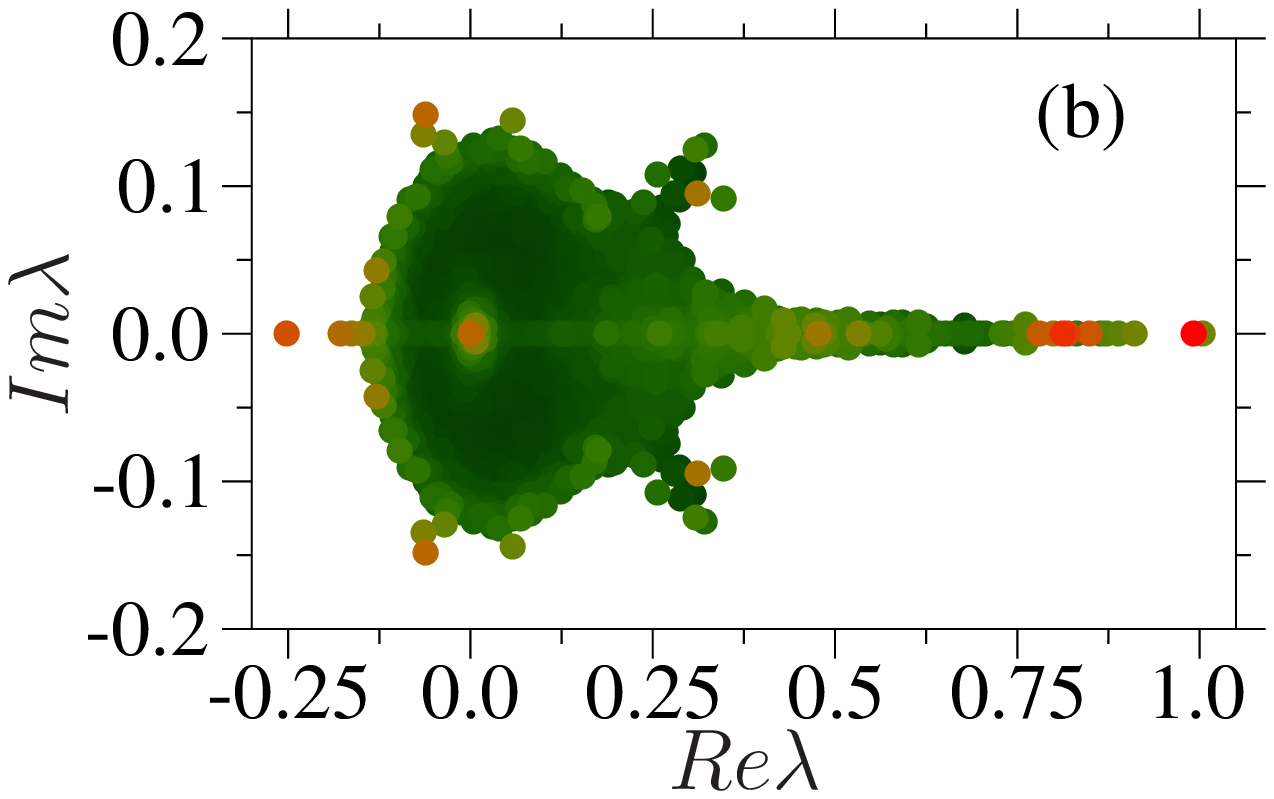}
}
\vglue -1.2cm
\centerline{
\epsfxsize=4.5cm\epsffile{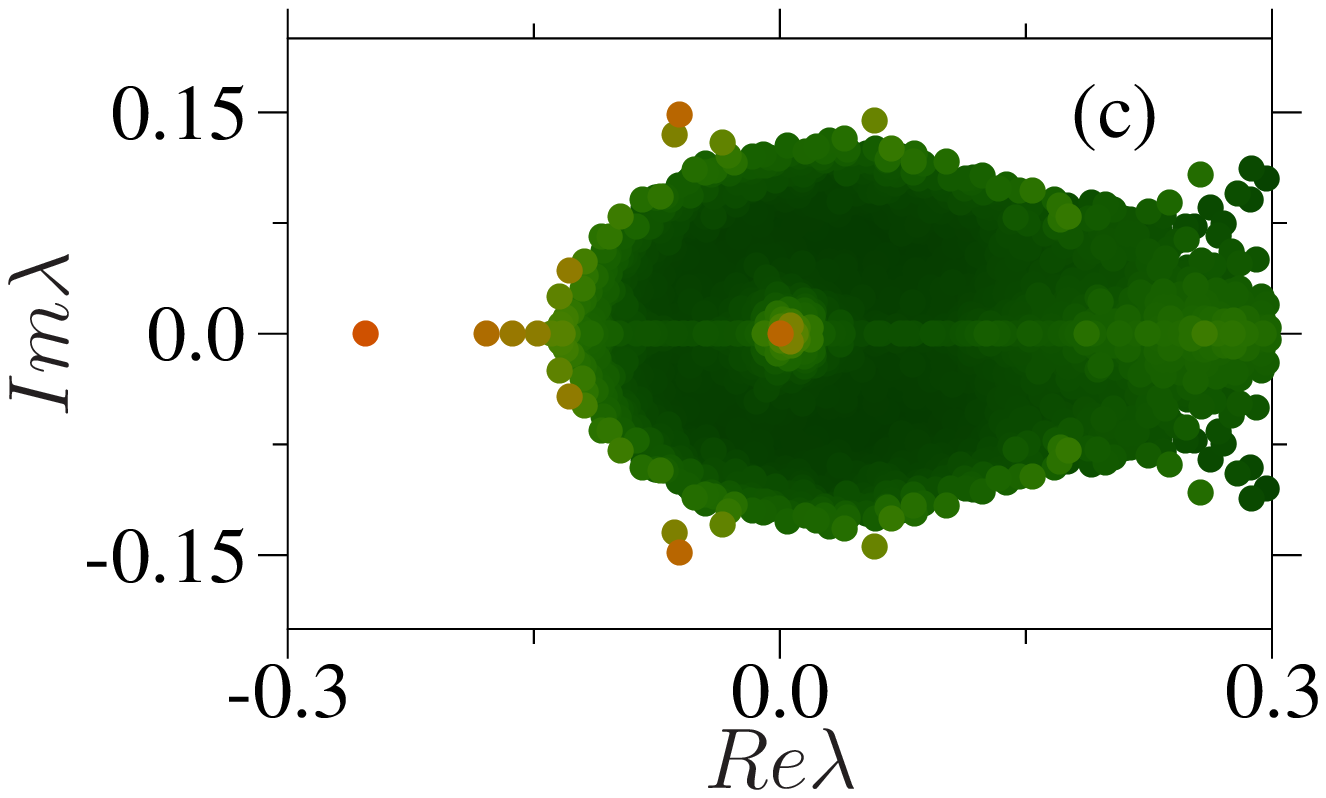}
\epsfxsize=4.5cm\epsffile{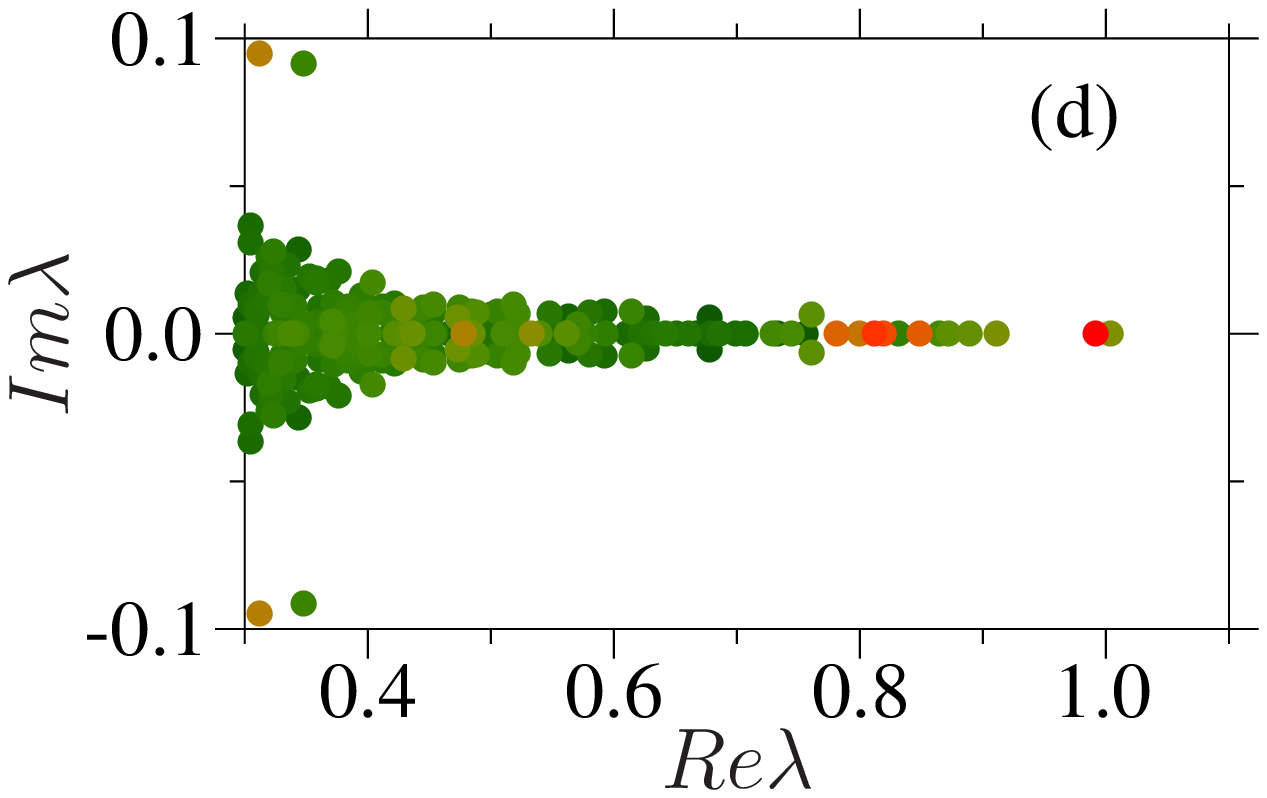}
}
\vglue -0.3cm
\caption{(Color online) Spectrum of eigenvalues  of the Google matrix
${\bf G}$ of
brain at $\alpha=0.99$ in the complex plain $\lambda$ for 
(a) unweighted and (b) weighted links in the neuronal network.
Panels (c) and (d) show zooms of data of panel (c).
The color shows
the degree of localization of eigenvectors of ${\bf G}$ 
being proportional to the value of PAR $\xi$
and changing from one (red/light gray) to maximal value
(dark green/black).
} 
\label{fig3}
\end{figure}

The spectrum $\lambda_i$ and the right eigenvectors 
$\psi_i$ of the Google matrix of
brain are defined by the equation
\begin{equation}
{\bf G} \psi_i = \lambda_i \psi_i \; .
\label{eq2}
\end{equation}
The spectrum of $\lambda$ is complex and is shown in Fig.~\ref{fig3}.
The color of points is chosen to be proportional to the
PArticipation Ratio (PAR)
defined as $\xi = (\sum_j |\psi_i(j)|^2)^2/\sum_j|\psi_i(j)|^4$.
This quantity determines an effective number of sites
populated by an eigenstate $\psi_i$, it is often used 
to characterize localization-delocalization transition in
quantum solid-state systems with disorder (see e.g. \cite{mirlin}).
The spectrum has eigenvalues with $|\lambda_i|$ being close to unity
so that there is no gap in the spectrum of $\lambda$ in the vicinity of
$\lambda=1$ (we remind that the second 
term in the r.h.s. of (\ref{eq1}) transfers
$\lambda_i$ to $\alpha \lambda_i$ keeping only
one $\lambda_1=1$ \cite{googlebook}). 
This is different from the spectrum of 
random scale-free networks characterized by a large gap in the
spectrum of $\lambda$ \cite{ggs}.
 
\begin{figure}
\centerline{
\epsfxsize=7.5cm\epsffile{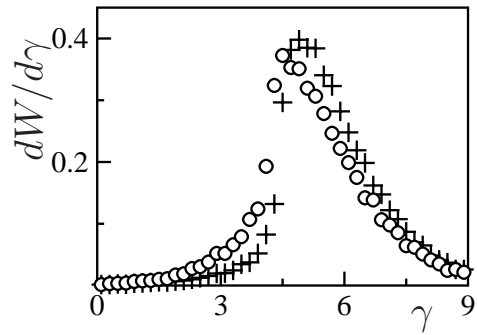}
}
\vglue -0.3cm
\caption{Dependence of the density of 
states $dW/d\gamma$ of ${\bf G}$ on the relaxation rate $\gamma$ 
for  unweighted (pluses) and weighted (circles) links in the neuronal network.
} 
\label{fig4}
\end{figure}

Compared to the spectra of the university WWW networks
studied in \cite{ggs2010} the spectrum of ${\bf G}$ in Fig.~\ref{fig3}
is more flat being significantly compressed to the real axis.
In this respect our neuronal network  has certain similarity
with the spectra of vocabulary networks 
analyzed in \cite{ggs2010} (see Fig.~1 there).
At the same time the spectrum of ${\bf G}$ matrix of brain
has visible structures in the eigenvalues distribution
in the complex plane of $\lambda$ while
the vocabulary networks are characterized by
structureless spectrum. 
The spectrum of Fig.~\ref{fig3}
has global properties being similar to those of the Ulam
networks considered in \cite{sz}. It is interesting to note that
the spectra of unweighted and weighted networks of brain
have similar structure. This supports the view of 
structural stability of the spectrum of ${\bf G}$ matrix.

It is useful to determine the relaxation rate
of eigestates by the relation $\gamma= - 2 \ln |\lambda|$.
The dependence of density of states $d W/d\gamma$ on $\gamma$ 
is shown in Fig.~\ref{fig4} (the density is normalized to 
unity so that $\int_0^{\infty} d W/d \gamma d \gamma =1$
corresponds to $N=10^4$ states). The distribution in $\gamma$
has a pronounced peak at $\gamma \approx 5$,
the density of states at small $\gamma<1$ is relatively small
(this is also seeing in Fig.~\ref{fig3}).
The comparison of unweighted and weighted links 
shows the stability of the density distribution
in respect to such modification of links.
\begin{figure}
\centerline{
\epsfxsize=4.5cm\epsffile{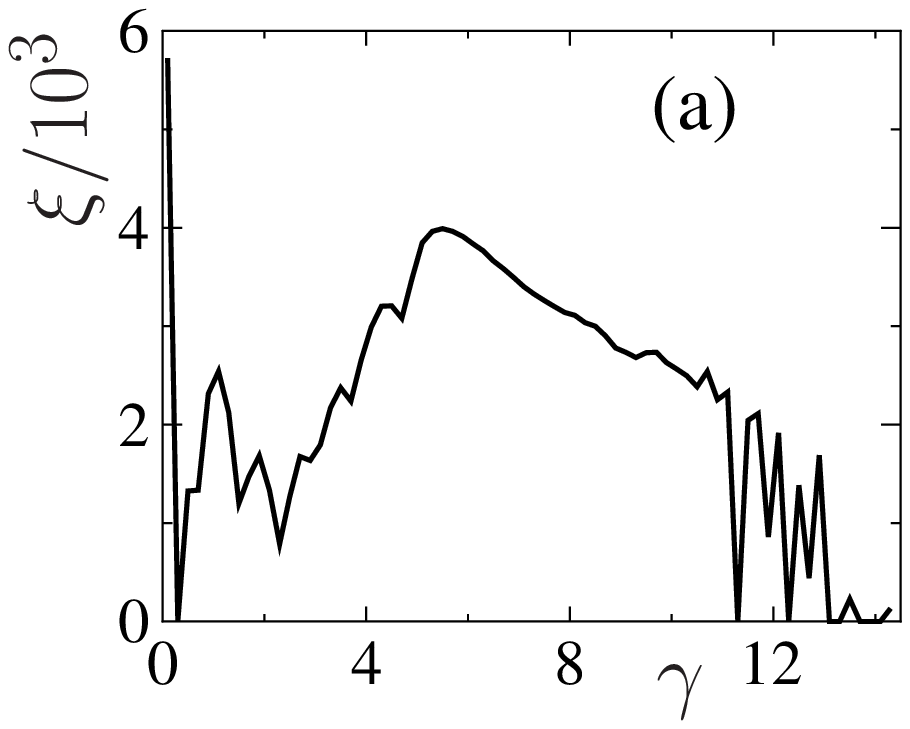}
\epsfxsize=4.5cm\epsffile{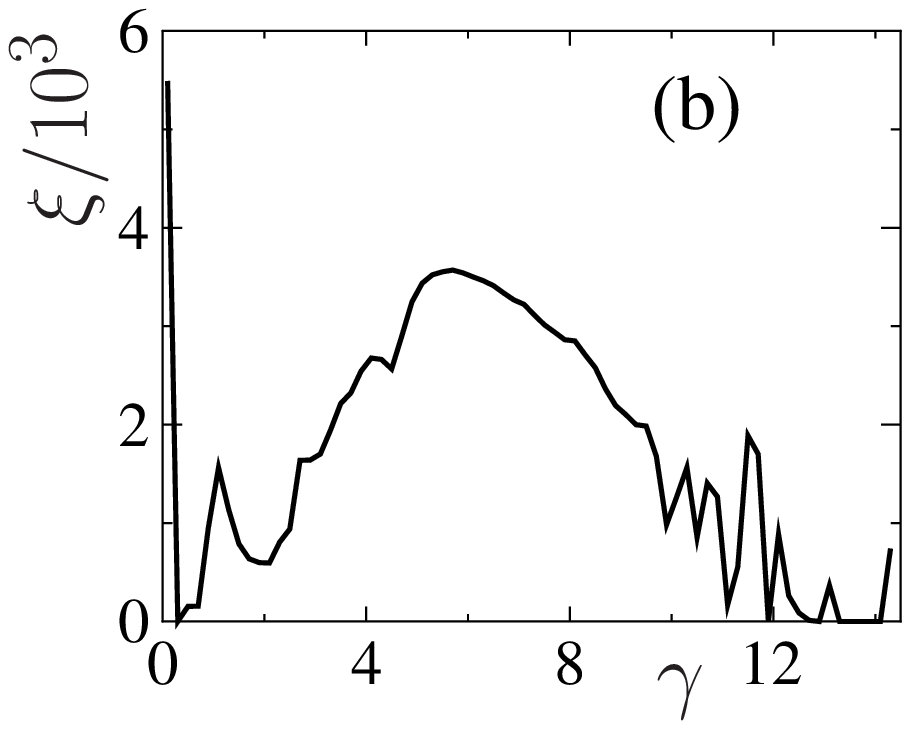}
}
\vglue -0.3cm
\caption{Dependence of 
PAR $\xi$ on relaxation rate $\gamma$ at $\alpha=0.85$
for (a) unweighted and (b) weighted links in the neuronal network.
} 
\label{fig5}
\end{figure}

\begin{figure}
\centerline{
\epsfxsize=4.5cm\epsffile{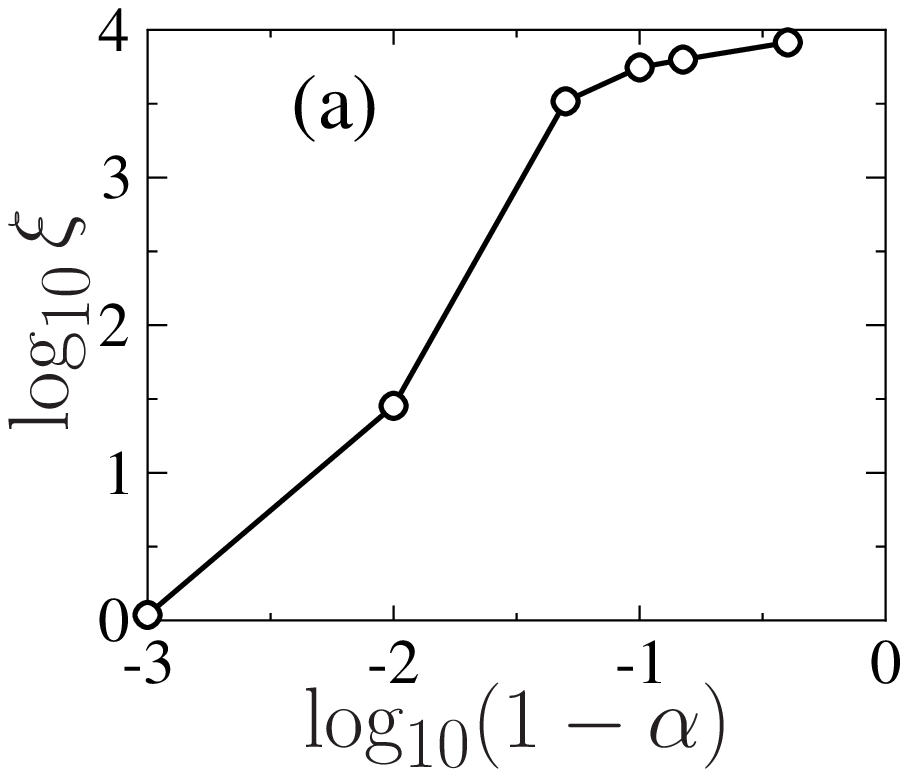}
\epsfxsize=4.5cm\epsffile{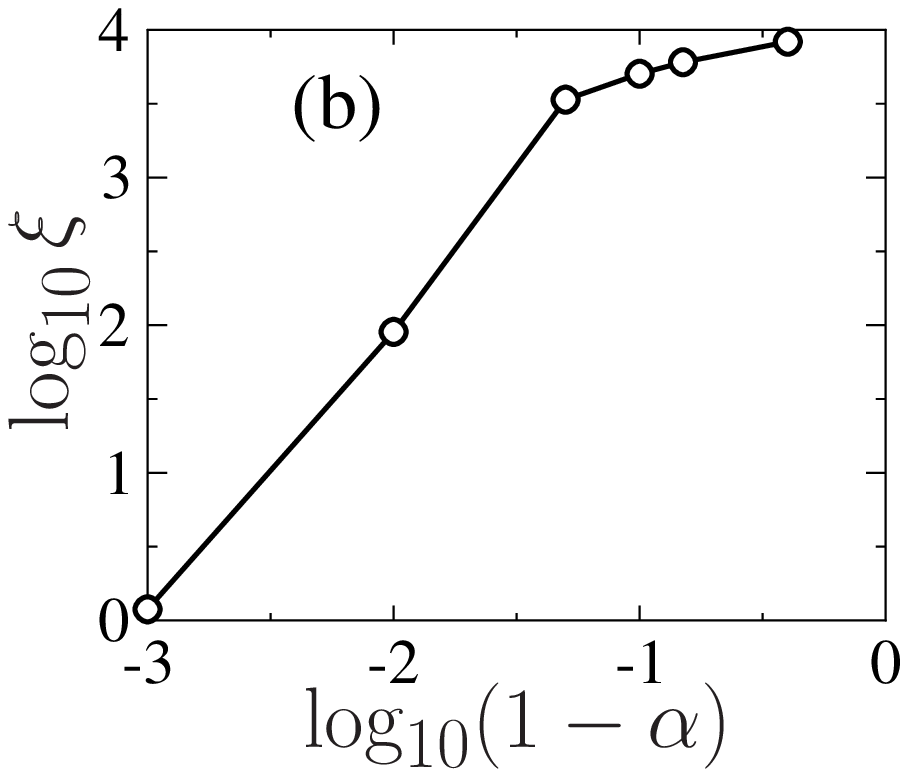}
}
\vglue -0.3cm
\caption{Dependence of 
PAR $\xi$ of the PageRank on parameter $\alpha$
for (a) unweighted and (b) weighted links in the neuronal network.
} 
\label{fig6}
\end{figure}

The dependence of the PAR $\xi$ on $\gamma$ is shown in Fig.~\ref{fig5}
(we note that except of the PageRank $\xi$ is independent of $\alpha$
due to the unity rank of matrix   ${\bf E}$, see e.g. \cite{googlebook,sz}).
The PageRank value of $\xi$ at $\gamma=0$  is very large
being more than half of the total number of neurons $N=10^4$.
It is clear that this corresponds to a delocalized state.
The eigenstates with $0<\gamma < 2$ have relatively small
$\xi \lesssim 10^3$ being close to a localized domain
while eigenstates with $2 < \gamma < 10$ have $\xi >  10^3 $
being delocalized on the main part of the network;
the states with $\gamma > 10$ enter in the localized domain.
For $\alpha > 0.99$ the PAR is close to $\xi \approx 1$.
Taking as a criterion that the delocalization takes place when
$\xi > N/2$ we obtain that 
the PageRank becomes delocalized at
$\alpha_c \approx 0.9$ (see data of Figs.~\ref{fig2},\ref{fig6}).
The global dependence of the PAR $\xi$ of the PageRank
on parameter $\alpha$ is shown in Fig.~\ref{fig6}
with a sharp delocalization of $\xi$ for $\alpha < \alpha_c$.
Of course, the above analysis should be considered 
as an approximate one
since the localization properties should be 
studied in dependence on the system size $N$
while we consider only one size of $N$.

\begin{figure}
\centerline{
\epsfxsize=4.5cm\epsffile{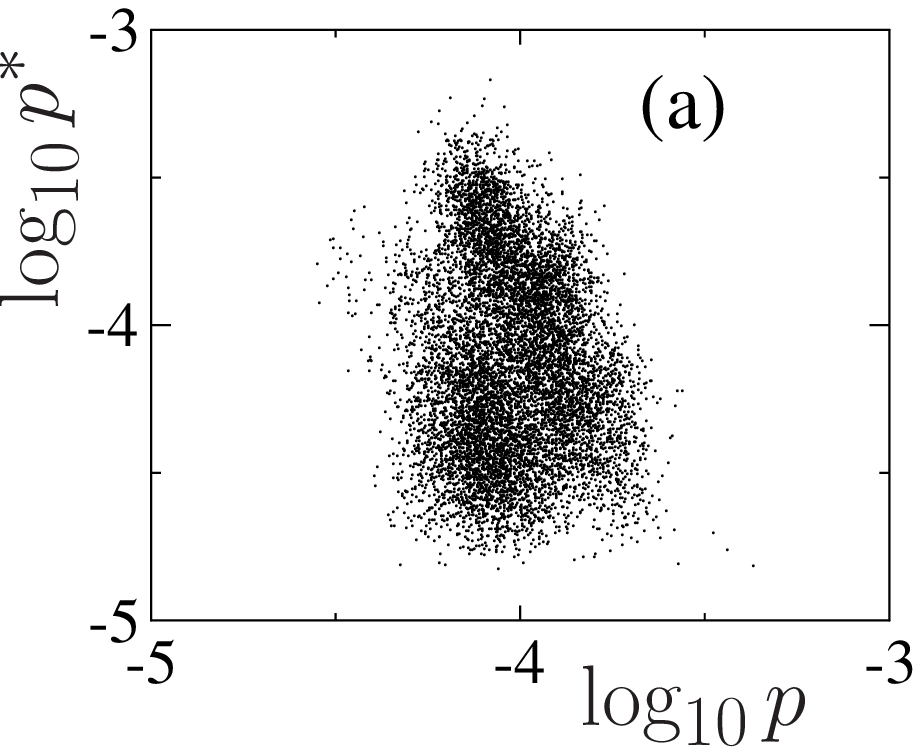}
\epsfxsize=4.5cm\epsffile{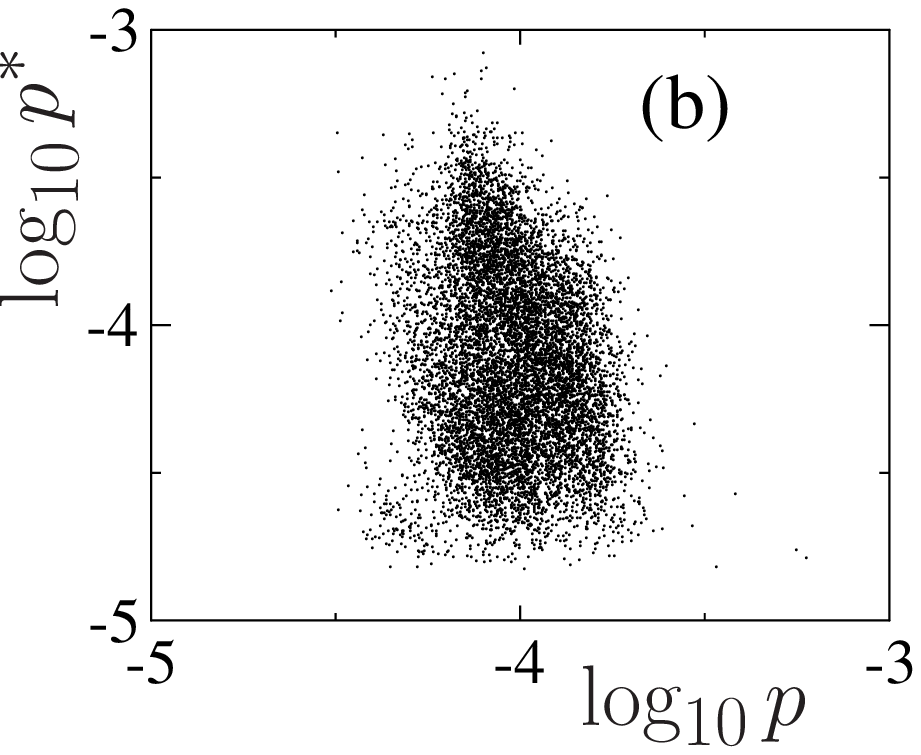}
}
\vglue -0.3cm
\caption{Distribution of PageRank values  $p$  and $p^*$
for all sites  
for (a) unweighted and (b) weighted links in the neuronal network
at $\alpha=0.85$.
} 
\label{fig7}
\end{figure}

Finally, following
the approach proposed in \cite{linux},
we show in Fig.~\ref{fig7} the distribution of PageRank
 values $p$  and $p^*$ for all sites. Such kind of distributions
can be rather useful in determining 
sites which have maximal values of $p$ and $p^*$
at the same time. However, a detailed analysis of the properties
of this distribution would require networks with a larger size $N$
where statistical fluctuations are smaller.

\section{III Discussion}

In this work we studied the properties of the Google matrix of 
a neuronal network of the brain model discussed in \cite{izhikevich3}.
For this network of $10^4$ neurons we found that the spectrum
of the Google matrix has a gapless spectrum at $\alpha=1$ demonstrating
certain similarities with  the spectra of university WWW networks
and vocabulary networks studied in \cite{ggs2010}. At the same time 
our neuronal network shows signs of delocalization transition
of the PageRank 
at the Google damping factor $\alpha_c \approx 0.9$
which was absent in the networks studied in \cite{avrach3,ggs2010}.
A similar transition in $\alpha$ was detected in the Ulam networks
generated by dissipative dynamical maps \cite{sz}.
We attribute the appearance of such delocalization transition
to a large number of links per neuron (200) which is
by factor 10 larger than in the WWW networks (20).

Of course, our studies have certain limitations 
since we considered only a fixed size neuronal network
and since this network is taken from a model system of
brain analysed in \cite{izhikevich3}.
Another weak point is that we do not consider
the dynamical properties of the network 
which are probably more important for practical applications.
Nevertheless, the spectral properties of ${\bf G}$ matrix
can be rather useful. Indeed, the gapless spectrum of $\lambda$
shows that long living excitations can exist
in our neuronal network. Such relaxation modes with small
rates $\gamma$ can be the origin
of long living oscillations found in numerical
simulations \cite{izhikevich3}. It is quite possible that
the properties of spectra of ${\bf G}$
can help to understand in a better way rapid relaxation processes
and those with long relaxation times.
We conjecture that the rapid relaxation modes 
correspond to relaxation of local groups of neurons
while long living modes can represent relaxation of collective
modes representing dynamics of human thoughts.
The dynamics of such collective modes can contain
significant elements of chaotic dynamics 
as it was discussed in the frame of the concept 
of creating chaos in \cite{chirikov}.

It is possible that the brain effectively implements dynamics
described by the evolution equation
$d \psi /d t = {\bf G} \psi$ which
without perturbations converges to the steady-state
described by the PageRank (which may be linked with 
a sleeping phase). External perturbations
give excitations of other eigenmodes of ${\bf G}$
discussed here. The evolution of these excitations will 
be significantly affected by the spectrum of ${\bf G}$.  

Further development of the Google matrix approach 
to the brain looks to us to be rather promising.
For example, a detection of isolated communities
and personalized PageRank, represented by 
other types of matrix ${\bf E}$ in (\ref{eq1}), 
is under active investigation in the computer science community
(see e.g. \cite{googlebook,avrach3}).
Such type of problems can find their applications
for detection of specific quasi-isolated
neuronal networks of brain.
The usage of real neuronal networks, similar to those
studied in \cite{laughlin,sporns2,sporns3,kaiser,sporns4,monasson},
in combination with the Google matrix approach
can allow to  discover new properties of processes in the brain.
The development of parallels between the WWW and neuronal networks
will give new progress of the ideas of John von Neumann.

We thank E.M.~Izhikevich for providing us with
the data set of links between neurons \cite{izhikevich4}
in the brain model \cite{izhikevich3}.


\begin{thebibliography}{99}
\bibitem{neumann} J. von Neumann, {\it The computer and the brain},
        New Haven, CT, Yale Uiv. Press (1958).
\bibitem{izhikevich1} F.C.~Hoppensteadt and E.M.~Izhikevich,
        {\it Weakly connected neural networks},
         Springer-Verlag, N.Y. Inc. (1997).
\bibitem{izhikevich2} E.M.~Izhikevich, 
        {\it Dynamical Systems in Neuroscience: 
        The Geometry of Excitability and Bursting}, 
        The MIT Press, Cambridge, MA (2007).
\bibitem{sporns1} O.~Sporns, {\it Brain connectivity}, 
        Scholarpedia { 2}(10) (2007) 4695.
\bibitem{felleman} D.J.~Felleman and D.C. van Essen, 
        Cereb. Cortex { 1} (1991) 1.
\bibitem{laughlin} S.B.~Laughlin and T.J.~Sejnowski,
        Science { 301} (2003) 1870.
\bibitem{sporns2} O.~Sporns, D.R.~Chialvo, M.~Kaiser, and C.C.~Hilgetag,
        TRENDS Cognitive Sci. { 8}, (2004) 418.
\bibitem{sporns3} C.J.~Honey, R.~K\"otter, M.~Breakspear, and O.Sporns,
        PNAS { 104} (2007) 10240.
\bibitem{kaiser} M.~Kaiser,
        Phil. Trans. R. Soc. A { 365} (2007) 3033.
\bibitem{sporns4} P.~Hagmann, L.~Cammoun, X.~Gigandet, R.~Meuli, C.J.~Honey,
        V.J.~Weeden, and O.Sporns,
         PLOS Biology { 6} (2008)  1479.
\bibitem{izhikevich3} E.M.~Izhikevich and G.M.~Edelman,
        PNAS { 105} (2008) 3593. 
\bibitem{chklov} Q.~Wen, A.~Stepanyants, G.N.~Elston, A.Y.~Grosberg,
        and D.B.~Chklovskii,
        PNAS { 106} (2009) 12536. 
\bibitem{monasson} S.~Cocco, S.~Leibler, and R.~Monasson,
        PNAS {106} (2009) 14058.     
\bibitem{brin} S.~Brin and L.~Page, Computer Networks and ISDN Systems {\bf 30} (1998)
             107.
\bibitem{googlebook} A.~M.~Langville and C.~D.~Meyer, {\it Google's
                      PageRank and Beyond: The Science of 
                      Search Engine Rankings},
                      Princeton University Press (Princeton, 2006); D.~Austin,
                       AMS Feature Columns (2008) available at
                       www.ams.org/featurecolumn/archive/pagerank.html
\bibitem{sinai} I.P.~Cornfeld, S.V.~Fomin, and Y.~G.~Sinai,
                {\it Ergodic theory}, Springer, N.Y. (1982).
\bibitem{mbrin} M.~Brin and G.~Stuck, 
        {\it Introduction to dynamical systems},
        Cambridge Univ. Press, Cambridge, UK (2002).
\bibitem{donato} D.~Donato, L.~Laura, S.~Leonardi and 
              S.~Millozzi, Eur. Phys. J. B {38} (2004) 239;
                  G.~Pandurangan, P.~Raghavan and E.~Upfal, Internet
                  Math. { 3} (2005) 1.
\bibitem{boldi} P.~Boldi, M.~Santini, and S.~Vigna, in
        {\it   Proceedings of the 14th international conference 
        on World Wide Web},
        A.~Ellis and T.~Hagino (Eds.), ACM Press, New York p.557 (2005);
        S.~Vigna, {\bf ibid.} p.976.
\bibitem{avrach1} K. Avrachenkov and D. Lebedev, 
        Internet Math. { 3} (2006) 207.
\bibitem{avrach2}  K.~Avrachenkov, N.~Litvak, and K.S.~Pham,
        in {\it Algorithms and Models for the Web-Graph: 
        5th International Workshop, 
        WAW 2007 San Diego, CA, Proceedings},
        A.~Bonato and F.R.K.~Chung (Eds.), Springer-Verlag, Berlin,
        Lecture Notes Computer Sci. { 4863} (2007) 16.
\bibitem{litvak} N. Litvak, W. R. W. Scheinhardt, and Y. Volkovich,
        Internet Math. { 4} (2007) 175.
\bibitem{avrach3} K.~Avrachenkov, D.~Donato and N.~Litvak (Eds.),
        {\it Algorithms and Models for the Web-Graph: 
        6th International Workshop, 
        WAW 2009 Barcelona,  Proceedings},  Springer-Verlag, Berlin,
        Lecture Notes Computer Sci. { 5427}, Springer, Berlin (2009).
\bibitem{fortunato1} N.~Perra and S.~Fortunato,
         Phys. Rev. E { 78} (2008) 036107.
\bibitem{dorogovtsev} S.N.~Dorogovtsev and J.F.F.~Mendes,
        {\it Evolution of Networks}, Oxford Univ. Press, Oxford (2003).
\bibitem{redner} P.~Chen, H.~Xie, S.~Maslov, and S.~Redner,
        J. Infometrics { 1} (2007) 8.
\bibitem{fortunato2}  F.~Radicchi, S.~Fortunato, 
        B.~Markines, and A.~Vespignani,
        Phys. Rev. E {\bf 80} (2009) 056103.
\bibitem{izhikevich4} E.M~Izhikevich, private communication, August (2009):
        the links between neurons
        have been generated by 
        E.M.~Izhikevich on the basis of the brain model 
        described in \cite{izhikevich3}; the links are available at
        Quantware Library, K.~Frahm and D.Shepelyansky (Eds.)
        section QNR15 at www.quantware.ups-tlse.fr/QWLIB
\bibitem{sz} D.L.~Shepelyansky and O.V.Zhirov, 
        Phys. Rev. E  81 (2010) 036213.
\bibitem{ermann} L.~Ermann and D.L.~Shepelyansky,
        Phys. Rev. E  81 (2010) 036221.
\bibitem{ggs2010} B.~Georgeot, O.~Giraud and D.L.~Shepelyansky,
         preprint arXiv:1002.3342[cs.IR] (2010).
\bibitem{linux} A.D.~Chepelianskii, arXiv:1003.5455[cs.SE] (2010).
\bibitem{mirlin} F.~Evers and A.D.~Mirlin, 
        Rev. Mod. Phys. {80} (2008) 1355.
\bibitem{ggs} O.~Giraud, B.~Georgeot and D.~L.~Shepelyansky, 
              Phys. Rev. E { 80} (2009) 026107.
\bibitem{chirikov} B.V.Chirikov, {\it Creating chaos and the Life}, 
        preprint arXiv:physics/0503072 (2005).


\end{thebibliography}
\end{document}